\documentclass{ecai} 

\usepackage{cite}
\usepackage{hhline}
\usepackage[numbers]{natbib}
\usepackage[utf8]{inputenc}
\usepackage{graphicx}

\usepackage{multirow}
\usepackage{threeparttable}
\usepackage{eurosym,makecell,comment}
\usepackage{url}
\usepackage{xcolor}
\usepackage{booktabs}
\usepackage{float}
\usepackage{algorithm}
\usepackage[noend]{algpseudocode}
\usepackage{amsmath,amssymb,amsfonts}
\usepackage[none]{hyphenat}
\usepackage{tabularx}
\usepackage{tabularray}
\usepackage[font=small,skip=5pt]{caption}
\usepackage{array}
\newcolumntype{P}[1]{>{\raggedright\arraybackslash}p{#1}}
\usepackage{dblfloatfix}
\usepackage{subfig}

\def \1{\textit{(i)}}
\def \2{\textit{(ii)}}
\def \3{\textit{(iii)}}
\def \4{\textit{(iv)}}
\def \5{\textit{(v)}}

\begin{document}


\begin{frontmatter}

\paperid{123} 

\title{Sentinel: An Aggregation Function to Secure Decentralized Federated Learning}

\author[A]{\fnms{Chao}~\snm{Feng}\thanks{Corresponding Author. Email: cfeng@uzh.ch.}}
\author[A]{\fnms{Alberto}~\snm{Huertas Celdr\'an}}
\author[A]{\fnms{Janosch}~\snm{Baltensperger}} 
\author[B]{\fnms{Enrique Tom\'as}~\snm{Mart\'inez Beltr\'an}} 
\author[B]{\fnms{Pedro Miguel}~\snm{S\'anchez S\'anchez}} 
\author[C]{\fnms{G\'er\^ome}~\snm{Bovet}} 
\author[A]{\fnms{Burkhard}~\snm{Stiller}}

\address[A]{Communication Systems Group CSG, Department of Informatics, University of Zurich UZH, CH--8050 Zürich, Switzerland}
\address[B]{Department of Information and Communications Engineering, University of Murcia, 30100--Murcia}
\address[C]{Cyber-Defence Campus, armasuisse Science \& Technology, CH--3602 Thun, Switzerland}

\begin{abstract}


Decentralized Federated Learning (DFL) emerges as an innovative paradigm to train collaborative models, addressing the single point of failure limitation. However, the security and trustworthiness of FL and DFL are compromised by poisoning attacks, negatively impacting its performance. Existing defense mechanisms have been designed for centralized FL and they do not adequately exploit the particularities of DFL. Thus, this work introduces Sentinel, a defense strategy to counteract poisoning attacks in DFL. Sentinel leverages the accessibility of local data and defines a three-step aggregation protocol consisting of similarity filtering, bootstrap validation, and normalization to safeguard against malicious model updates. Sentinel has been evaluated with diverse datasets and data distributions. Besides, various poisoning attack types and threat levels have been verified. The results improve the state-of-the-art performance against both untargeted and targeted poisoning attacks when data follows an IID (Independent and Identically Distributed) configuration. Besides, under non-IID configuration, it is analyzed how performance degrades both for Sentinel and other state-of-the-art robust aggregation methods.

\end{abstract}

\end{frontmatter}

\section{Introduction}

Federated Learning (FL) emerged in 2016 as a promising paradigm able to train Machine and Deep Learning (ML \& DL) models in a privacy-preserving and collaborative manner \citep{alazab_FederatedLearning_2022}. Nowadays, FL is used in many different scenarios such as healthcare, cybersecurity, networking, and so on. More in detail, FL combines the capabilities of communication systems with ML to solve data privacy concerns and increase scalability while distributing the learning process throughout a federation of participants. However, the original concept of FL relies on a central entity aggregating weights to create global models, which implicates drawbacks such as communication bottlenecks and single points of failure. In this context, Decentralized FL (DFL) has been recently proposed as an alternative to improve these limitations \citep{Matinez_DFLSurvey_2023}. In DFL, communication systems get much more relevance than in CFL, since the aggregation task is decentralized and different network topologies, communication protocols, and network optimization techniques play a key role. 

Despite the merits of both centralized and decentralized FL, recent research has unveiled their vulnerabilities to adversarial attacks, particularly poisoning attacks that manipulate training data \citep{rodriguez-barroso_SurveyFederated_2023}. FL is highly susceptible to such attacks because traditional data inspection techniques cannot be used to effectively detect altered data, creating a challenge in balancing privacy protection with trust and robustness establishment \citep{sanchez_2023_federatedtrust}. Successful poisoning in FL can yield significant repercussions, including inaccurate predictions, misclassification of inputs, and the introduction of exploitable backdoors within compromised models \citep{xia_PoisoningAttacks_2023}. Furthermore, these attacks can profoundly disrupt the collaborative learning process, eroding trust among participants and undermining the core privacy and security advantages of FL.

To defend against poisoning attacks in Centralized Federated Learning (CFL), researchers have actively explored various robust and secure techniques dealing with Byzantine-robust aggregation, anomaly detection, and hybrid approaches \citep{tian_ComprehensiveSurvey_2023}. These defenses primarily concentrate on detecting and mitigating the detrimental effects of poisoning attacks, striving to uphold the integrity of FL and sustain the performance of individual participants' models. Nevertheless, some challenges emerge in the transition to DFL, where defense techniques are unexplored. When comparing DFL with CFL, DFL has the advantage of leveraging more information during the aggregation process, such as local data, weights, and models (apart from network topology information), which could be used to enhance the defenses effectiveness. In this sense, by comparing the similarity between the local model and the shared models and calculating the loss of the shared models on the local dataset, the DFL participant could improve the security of the model aggregation. 

To improve the aforementioned limitations, this work introduces Sentinel, a hybrid defense strategy designed to enhance the resilience of DFL against poisoning attacks. Sentinel proposes a novel three-phase aggregation protocol to bolster security. In the initial phase, it employs a layer-wise average cosine similarity metric to identify and filter highly suspicious model updates. Subsequently, the remaining updates are aggregated based on local bootstrap validation loss, with aggregation weights determined through adaptive loss distance mapping. In the final phase, trusted models are normalized using the local model norm as a threshold to mitigate the impact of potential stealth attacks. Sentinel performance has been evaluated in a well-established DFL framework, Fedstellar~\citep{MartinezBeltran:fedstellar:2023}, utilizing Computer Version (CV) datasets: MNIST, Fashion-MNIST, CIFAR10, Natural Language Processing (NLP) dataset: Sentiment140, and tabular dataset: Purchase, following IID (Independent and Identically Distributed) and non-IID data distributions. It undergoes evaluation under various data and model poisoning attacks in both targeted and untargeted forms, with varying numbers of attackers modifying different data portions. Results show high performance on IID, improving other state-of-the-art methods, while non-IID results remark the drawbacks of extensive filtering, something common to other robust aggregation methods.

The remainder of this document is organized as follows. First, Section \ref{sec:related} reviews previously proposed defense mechanisms for CFL and DFL. Section~\ref{sec:sentinel} introduces the design details of Sentinel, the proposed robust aggregation mechanism. Then, Section~\ref{sec:experiments} evaluates and compares the performance of Sentinel with related work. Section~\ref{sec:discussion} provides the discussion of the achieved results. Lastly, Section \ref{sec:conclusion} summarizes the contributions of this work and proposes future opportunities.

\section{Related Work}
\label{sec:related}

The most prominent methods to mitigate poisoning attacks are designed for CFL, where the model aggregation is done by a central server to reduce complexity. As indicated in \citep{tian_ComprehensiveSurvey_2023}, existing mitigation approaches are categorized into: Byzantine-robust aggregation, anomaly detection, and hybrid approaches. 

\begin{table}[ht]\centering
\caption{Techniques Against Poisoning Attacks. U Stands for Untargeted Attacks and T Stands for Targeted Attacks}
\label{tab:related}
\resizebox{\columnwidth}{!}{%
\renewcommand{\arraystretch}{1.2}
\newcolumntype{L}[1]{>{\raggedright\let\newline\\\arraybackslash\hspace{0pt}}m{#1}}
\setlength\tabcolsep{1.5pt}
\begin{tabular}{cclL{1in}ccc}
\toprule
\multirow{2}{*}{\textbf{Category}} &\multirow{2}{*}{\textbf{Type}} &\multirow{2}{*}{\textbf{Method}} &\multirow{2}{*}{\textbf{Technique}} &\multirow{2}{*}{\textbf{Schema}} &\multicolumn{2}{c}{\textbf{Obj.}} \\\cmidrule(lr){6-7}
\textbf{} &\textbf{} &\textbf{} &\textbf{} & & U & T \\\midrule
    & & COMED \citep{yin_ReputationBasedResilient_2017} &Coordinate-wise median &CFL &\checkmark &x \\
    & &RFA \citep{pillutla_RobustAggregation_2022} &Geometric median &CFL &\checkmark &x \\
    Robust& Geometry &TrimmedMean \citep{yin_ByzantineRobustDistributed_2018} &Filtered mean &CFL &\checkmark &x \\
    Aggregation&  &Krum \citep{blanchard_MachineLearning_2017} &Euclidean distance &CFL &\checkmark &x \\
    & &Multi-Krum \citep{blanchard_MachineLearning_2017} &Euclidean distance &CFL &\checkmark &x \\
    & &Bulyan \citep{mhamdi2018hidden} &Krum and TrimmedMean &CFL &\checkmark &x \\ 
    \cmidrule{1-7}
    Anomaly &Validation &ERR, LFR \citep{fang_LocalModel_2021} &Global validation &CFL &\checkmark &\checkmark \\
    Detection & &PDGAN \citep{zhao2020pdgan} &Model accuracy auditing &CFL &\checkmark &\checkmark \\ \cmidrule(l){2-7}
    & Gradient &FLDetector \citep{zhang_FLDetectorDefending_2022} &Hessian-based gradient consistency &CFL &\checkmark &\checkmark \\
    \cmidrule{1-7}
     &  &FLTrust \citep{cao_FLTrustByzantinerobust_2021} &ReLU-clipped cosine similarity, norm thresholding &CFL &\checkmark &\checkmark \\
    Hybrid &  &Gholami et al. \citep{gholami_TrustedDecentralized_2022} &Trusted aggregation &DFL &\checkmark &x \\
    & &FLAME \citep{nguyen_FLAMETaming_2021} &Clustering (cosine similarity), adaptive clipping, noising &CFL &\checkmark &\checkmark \\\cmidrule{1-7}
    This work & & Sentinel & Model similarity, bootstrap validation and normalization & DFL & \checkmark & \checkmark \\
    \bottomrule
    \end{tabular}    
}
\end{table}

Byzantine-robust aggregation mechanisms try to prevent malicious updates deteriorating the model performance. Among the existing alternatives, Coordinate-wise Median is a defense method that applies dimension-wise filtering \citep{yin_ReputationBasedResilient_2017}. This approach is a generalization of the median in higher dimensions, is insensitive to skewed distributions, and effectively defends against a model replacement attack. TrimmedMean \citep{yin_ByzantineRobustDistributed_2018} is similar to the coordinate-wise median but excludes the lowest and largest values in each dimension of the sorted model updates. RFA \citep{pillutla_RobustAggregation_2022} is a simple alteration of FedAvg, where the mean aggregation is replaced with the geometric median. Krum \citep{blanchard_MachineLearning_2017} is a more sophisticated approach that assigns a score to each model update based on the similarity to other updates. Client scores are calculated by summing the Euclidean distance to the closest client updates. Bulyan \citep{mhamdi2018hidden} was proposed to address the limitations of Krum concerning the curse of dimensionality. However, experiments have demonstrated that in some scenarios, Bulyan performs worse than Krum due to potentially trimming benign updates.

Defense mechanisms through anomaly detection, also referred to as Byzantine detection, aim at identifying and removing potentially malicious updates. In contrast to Byzantine robustness, these anomaly detection schemes do not implement the defense strategy into the aggregation rules. \citep{fang_LocalModel_2021} proposed two adapted defense strategies for CFL: Error Rate based Rejection (ERR) and Loss Function based Rejection (LFR). The key concept of these strategies is to evaluate the performance of the collected client models on a server-side validation dataset. Before averaging the received models, a predefined number of updates that have the largest impact on either the loss or the validation error are rejected. As a drawback, the server is required to collect a clean data set, which may violate the privacy-preserving concept of FL. To overcome that limitation, the authors of \citep{zhao2020pdgan} proposed PDGAN. Instead of requiring a clean validation dataset, the system trains a generative adversarial network (GAN) concurrently to the original federated learning task. Once trained, PDGAN reconstructs a client's data from their model updates received on the server side. The generated data can be used to audit individual updates and label malicious clients as attackers. However, PDGAN can only defend from attackers after a certain number of iterations when the GAN is trained. Another approach is to investigate the gradient consistency of client updates. In this direction, FLDetector \citep{zhang_FLDetectorDefending_2022} predicts a client's update based on past contributions. Specifically, the server applies a quasi-newton approach to estimate the Hessian of an update and compare it to the integrated Hessian. FLDetector is effective against various adaptive attacks, such as distributed backdoors, and untargeted model poisoning. However, predicting consistency is computationally expensive.

Finally, hybrid defenses employ a combination of robust aggregation and anomaly detection mechanisms. \citep{cao_FLTrustByzantinerobust_2021} proposed FLTrust, a trusted aggregation mechanism for CFL based on comparing local model updates to a trusted bootstrapping model. Therefore, the server acquires a clean root dataset to train a reference model, and in each iteration, the received model updates are compared to the reference model. The result is then used as a trust score, and updates that deviate too much from the reference model receive a score of zero. \citep{gholami_TrustedDecentralized_2022} implemented the concept of trusted aggregation in the decentralized architecture. Each node is judged based on a behavioral score reflecting a participant's performance contribution and consistency. Computationally, this is represented by cluster-based and distance-based metrics. Each neighboring node's local trust score is then broadcast so that each node can compute a global trust score based on their neighbors' opinions. Finally, \citep{nguyen_FLAMETaming_2021} proposed FLAME, a hybrid approach based on dynamic clustering, adaptive clipping, and noising. In the first step, the client updates are clustered based on pair-wise cosine similarity to capture angular deviation. The authors argue that existing clustering-based approaches often group the models into malicious and benign. This leads to the issue that benign updates become removed when no adversaries are present. 

In conclusion, as can be seen in \tablename~\ref{tab:related}, a noticeable research gap exists concerning defense mechanisms in the domain of DFL. Previous attempts prominently explored defenses against poisoning attacks in CFL, but only a few works investigated defense mechanisms in DFL. Therefore, while DFL has gained significant attention as a promising approach without a central entity, the security aspects of poisoning attacks still need to be addressed.

\section{Sentinel}
\label{sec:sentinel}

\begin{figure}[htpb!]
\centering
\includegraphics[width=1\linewidth]{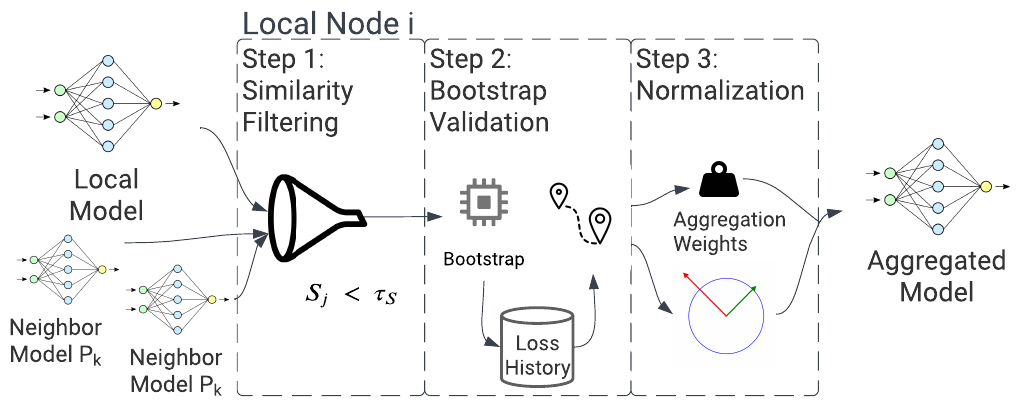}
\caption{Overview of the Sentinel Aggregation Process.}
\label{fig:sentinel_visual}
\end{figure}

Sentinel is a defense mechanism that aims to defend against poisoning attacks in DFL. It considers local data availability and relies on the cosine similarity and bootstrap loss to evaluate the trust performance of neighbors' models. As can be seen in \figurename~\ref{fig:sentinel_visual}, Sentinel is a three-phase aggregation protocol consisting of similarity filtering, bootstrap validation, and layer normalization. 

The formal overview of Sentinel is given in Algorithm \ref{alg:sentinel}. The local model $M$ is defined as the model trained on the individual node available in any round. Additionally, a model received from an adjacent node $n_i$ is referred to as a neighbor model $P_i$. More information about the three steps of Sentinel are elucidated in the following paragraphs. 

\begin{algorithm}[h]
\caption{Sentinel Aggregation Algorithm}\label{alg:sentinel}
\begin{algorithmic}[1]
\small
\Require $P$: Neighbor parameters, $D_{bs}$: Bootstrap dataset, $H$: Loss history, $\tau_{S}$: Similarity threshold,
$\tau_{L}$: Loss distance threshold.
\State $r \gets$ current round
\State $M \gets$ local model
\State $w \gets 0$
\For{$j$ in $P$, $j \neq i$} \Comment{\textbf{(1) Similarity filtering}}
    \State $S_j \gets CosineSimilarity(P_j, M)$
    \If{$S_j < \tau_{S}$} Remove $P_j$ from $P$
    \EndIf
\EndFor
\State $H_{i}[r] \gets ComputeBootstrapLoss(M, D_{bs})$ \Comment{\textbf{(2) Bootstrap validation}}
\For{$j$ in $P$, $j \neq i$} 
    \State $l_j \gets ComputeBootstrapLoss(P_i, D_{bs}$)
    \State $H_{j}[r] \gets l_j$ \Comment{Update loss history}
    \State $w_j \gets MapLossDistance(H_i, H_j, \tau_{L})$
\EndFor
\For{$j$ in $N$, $j \neq i$} \Comment{\textbf{(3) Layer Normalization}}
    \State $\Tilde{P_j} \gets NormaliseModel(P_i, P_j)$
\EndFor
\State $\theta \gets \frac{1}{\sum_{j \in N} w_j}\left(\sum_{j \in N} w_j \Tilde{P_j}\right)$ \Comment{Aggregate} \label{alg:sentinel:agg}
\State \Return $\theta$
\end{algorithmic}
\end{algorithm}

\textbf{Step 1: Similarity Filtering.} At the first stage, Sentinel computes the layer-wise average cosine similarity between the current local model and all neighbor models received. This similarity is denoted as $S_j$. The computation of the cosine similarity is defined in Algorithm \ref{alg:cossim}. The cosine similarity is chosen due to its advantage over the Euclidean distance. In contrast to existing methods, Sentinel does not cluster or weigh models based on cosine similarity, but rather filters them instead. This step serves as a preliminary similarity evaluation to exclude potentially malicious models and reduce the computation of the second stage. The decision boundary is defined by $\tau_{S}$, such that all models exhibiting a lower cosine similarity to the local model than $\tau_{S}$ will be filtered.

\begin{algorithm}[h]
\caption{Cosine Similarity}\label{alg:cossim}
\begin{algorithmic}[1]
\small
\Require $P$: Neighbor parameters, $M$: Local model parameters
\Ensure $|M| = |P|$
\For{$l_m, l_p$ in $M$, $P$}
    \State $S_P \gets S_P + \phi(\frac{\mathbf{v_1} \cdot \mathbf{v_2}}{ \|\mathbf{v_1}\| \|\mathbf{v_2}\|})$ \Comment{row-wise average}
\EndFor
\State $S_P \gets S_P / |M|$ \Comment{layer-wise average}
\State \Return $S_P$
\end{algorithmic}
\end{algorithm}

\textbf{Step 2: Bootstrap Validation.}
At the second step, the remaining models are evaluated on the basis of their loss. To do so, each node holds a small bootstrap dataset $D_{bs}$, which is randomly sampled from the validation dataset of the same node. The size of this bootstrap dataset is set to a third of the validation dataset or at least 300 samples, for example $|D_{bs}| = max(|D_{val}|, 300)$. The number of collected samples required to effectively measure the loss has been investigated in related studies and is therefore not within the scope of this work \citep{cao_FLTrustByzantinerobust_2021}. In general, choosing the bootstrap dataset size is a trade-off between computational overhead and performance. Therefore, the chosen size should be suitable for the device specifications of each participating node. The computation of the bootstrap validation loss is defined in Algorithm \ref{alg:bsloss}. Note that the local node also computes the bootstrap validation loss for its own local model for a more accurate comparison.

\begin{algorithm}[h]
\caption{Compute Bootstrap Loss}\label{alg:bsloss}
\begin{algorithmic}[1]
\small
\Require $P$: Model parameters, $D_{bs}$: Bootstrap dataset
\For{each $(x, y)$ in $D_{\text{bs}}$} 
    \State $y_{pred} \gets$ Predict with P on $x$
    \State $L \gets \frac{1}{\left|D_{\text{bs}}\right|} \sum_{i=1}^{\left|D_{\text{bs}}\right|} l\left(y, y_{\text{pred}}\right)$ \Comment{Compute loss}
\EndFor
\State \Return $L$
\end{algorithmic}
\end{algorithm}

After computing the performance of each received model, Sentinel computes the average of the loss history up to the current aggregation round. This procedure simply takes all previously computed loss values into account. Thus, if the bootstrap loss was not computed in a previous round, since the model was filtered in step (1), there will be consequently fewer values to be averaged. Generally, the averaging serves for better computational stability. Finally, all average loss values are then compared to the local average bootstrap loss according to Algorithm \ref{alg:maploss}, which results in an aggregation weight $w_i$.

\begin{algorithm}[h]
\caption{Map Loss Distance}\label{alg:maploss}
\begin{algorithmic}[1]
\small
\State $\Bar{l_i} \gets \phi(H_i)$, $\Bar{l_j} \gets \phi(H_j)$
\State $\kappa \gets max(\Bar{l_i}, l_{min})^{-1}$ \Comment{Damping factor}
\State $\Bar{d_l} \gets max(\Bar{l_j} - \Bar{l_i}, 0)$
\State $w \gets \exp{(-k * \Bar{d_l})}$
\If{$w < \tau_L$}
    $w = 0$
\EndIf
\State \Return $w$
\end{algorithmic}
\end{algorithm}

Sentinel maps the average loss distance according to a damped decay function.
The damping factor $\kappa$ is defined as the inverse of the average local loss. In theory, the local loss could be 0. Hence, a minimum average loss $l_{min}$ must be defined for numerical stability, for example $l_{min}=0.001$. Consequently, the local model and any neighbor model that presents a loss lower than the local model will receive $w = 1$.
With this approach, Sentinel becomes more defensive as the average local loss decreases, which is expected during the FL process. However, if the local loss increases, which means the node is not learning, Sentinel becomes more exploratory.


\textbf{Step 3: Layer Normalization.}
Inspired by recent approaches \citep{ cao_FLTrustByzantinerobust_2021}, the last step of Sentinel is the normalization of models to defend against potential stealth attacks that were able to pass defense layers (1) and (2). This procedure is defined in Algorithm \ref{alg:norm}. The normalization reduces the magnitude of potentially scaled attacks, which are commonly used to introduce backdoors. Such that Sentinel does not rely on a threshold, the ratio of the local model norm and the neighbor model norm is used as a scaling factor $\rho$. Sentinel only decreases the norm of neighbor models, hence $\rho \leq 1$. Subsequently, the normalized models are aggregated according to Algorithm \ref{alg:sentinel} line \ref{alg:sentinel:agg}.

\begin{algorithm}[h]
\caption{Normalize Model}\label{alg:norm}
\begin{algorithmic}[1]
\small
\Require $P$: Neighbor parameters, $M$: local model parameters
\State $L \gets$ number of layers
\For{l in $|M|$}
    \State $\rho \gets min(1, \frac{\lVert P[l] \rVert}{\lVert M[l] \rVert})$
    \State $\Tilde{P}[l] \gets \rho * P[l]$
\EndFor
\State \Return $\Tilde{P}$
\end{algorithmic}
\end{algorithm}

\section{Deployment and Experiments}
\label{sec:experiments}

Fedstellar~\citep{MartinezBeltran:fedstellar:2023} was chosen as the underlying platform to deploy and test Sentinel in DFL scenarios. Fedstellar allows users to train FL models in a decentralized, semi-decentralized, and centralized manner. It deploys the desired scenario to each specified client and manages the network of participants in terms of node connectivity. Users can run FL scenarios on selected physical devices or in a containerized simulation on Docker. Fedstellar is fully extensible, allowing users to implement custom models, load new datasets, and define the preferred aggregation algorithms.

\subsection{Deployment Configuration}

In this work, the Sentinel aggregation method (see Algorithm \ref{alg:sentinel}) was implemented in Python. PyTorch Lightning was used to implement the DL models. Then, for testing the Sentinel performance, a set of experiments was executed with the following configurations:

\begin{itemize}
    \item The federation is composed of 10 fully connected nodes, virtualized with Dockers, and acting as trainers and aggregators.
    \item Three types of datasets are employed to conduct experiments in this work. \1 Commonly used CV datasets  MNIST~\citep{deng2012mnist}, Fashion-MNIST~\citep{xiao2017fashion}, and CIFAR10 \citep{krizhevsky2009learning}; \2 NLP sentiment analysis dataset Sentiment140~\citep{go2009twitter}; and \3 tabular dateset Purchase~\citep{purchase}.
    
    \item For the IID evaluation, data are equally distributed among all nodes. For MNIST and Fashion-MNIST, each node receives $6\,000$ training and $1\,000$ test samples; for CIFAR10, $5\,000$ training and $1\,000$ test samples; for Sentiment140, $8\,500$ training and $1\,500$ test samples; for Purchase, $7\,384$ training and $1\,300$ test samples;  and the validation datasets are $10\%$ of the training sets. 

    \item For the non-IID evaluation, the same datasets are distributed across the participants by following a Dirichlet function \citep{wang2020federated} with $\alpha=0.5$, and using the same ratio for training, validation, and testing.

    \item The chosen model topology for MNIST and Fashion-MNIST datasets is a Multi-Layer Perceptron (MLP) containing two fully-connected hidden layers with 256 and 128 neurons. 
    \item For CIFAR10, the model topology is a Convolutional Neural Network (CNN), namely SimpleMobileNet \citep{sinha2019thin}, which contains several convolutional layers followed by pooling layers and fully connected layers towards the end. The first layer uses a set of 32 filters with a kernel size of 3x3. This is followed by a max pooling layer to reduce the spatial dimensions. Then, five depthwise separable convolutional layers with different input and output channel sizes and strides. These layers help reduce the model computational cost while preserving expressive power. Finally, global pooling in the form of an adaptive average pooling layer is applied to reduce the spatial dimensions of the output feature map to a size of 1×1. This vector feeds into a dense layer of 512 units. 

    \item For Sentiment140, a bidirectional Long Short-Term Memory (LSTM) network is used for the model topology. In the preprocessing of the data, the original text data is firstly tokenized using the lexicon of the pre-trained word embedding of Glove's Twitter.27B \citep{pennington2014glove}, and then the tokens are transformed into a 100-dimensional word vector. In order to make the length of each input vector the same, all samples are pad to 64 dimensions, i.e., after preprocessing, all the text is transformed into a 64*100 tensor. In terms of the model topology, after the input layer, a bi-directional LSTM with a hidden layer of 256 is utilized, and finally, a fully connected layer with an output layer of 2 is connected.

    \item MLP is employed as a model for the Purchase dataset, which contains only a 256-neuron hidden layer.
    
    \item In all model topologies, \textit{ReLU} is employed as an activation function in the hidden layers and \textit{softmax} in the output layer. \textit{Adam} is employed as an optimizer with default hyperparameters.
    
    \item The duration of the learning process is configured to 10 rounds of 3 epochs. To make experiments reproducible, synchronous aggregation is implemented in Fedstellar.

    \item The federation network bandwidth is configured to 1 Mbps, the loss to 0\%, and the delay to 0 ms.
    
    \item The ratio of poisoned nodes (PNR) is established to 10\%, 50\%, and 80\%. Malicious nodes are randomly selected, and they launch model poisoning, label flipping (targeted and untargeted), and backdoor attacks.

    \item FedAvg, FLTrust, Krum, TrimmedMean, and Sentinel are implemented in Fedstellar, evaluated, and compared. FedAvg acts as a baseline, and the other methods act as defense mechanisms based on robust aggregation.

\end{itemize}

\subsection{Experiments}

To compare the performance of Sentinel with existing defense mechanisms, a baseline performance reference under benign settings is established. Subsequently, a set of experiments evaluates the performance of each defense for the previous datasets and attacks.

\textbf{Baseline.} The goal of this experiment is to establish a baseline performance that serves as a reference for subsequent experiments in malicious environments. The evaluation of baseline, model poisoning, and untargeted label flipping attacks employs the F1-Score.


As can be seen in \tablename~\ref{tab:baseline-f1}, Sentinel achieves a similar F1-Score to related work for the five datasets when the dataset follows an IID distribution. In the case of non-IID, the F1-Score performance degrades compared to FedAvg in $\approx$0.05, 0.10, and 0.15 for MNIST, Fashion-MNIST and CIFAR10, and a decrement of 0.10 for both Sentiment140 and Purchase. However, this decrease is consistent with other robust aggregation mechanisms such as Krum. These results demonstrate that Sentinel is aligned with other robust mechanisms and is a valid option when no attacks affect the federation nodes.

\begin{table}[h]
\centering

\caption{F1-Score for MNIST, Fashion-MNIST, CIFAR10, Sentiment140, and Purchase without Attacks}
\label{tab:baseline-f1}
\resizebox{\columnwidth}{!}{%
\begin{tabular}{llllll}
\toprule
               & \multicolumn{1}{c}{MNIST} & \multicolumn{1}{c}{Fashion-MNIST} & \multicolumn{1}{c}{CIFAR10} & 
               \multicolumn{1}{c}{Sentiment140} &
               \multicolumn{1}{c}{Purchase}\\
\toprule
\multicolumn{6}{c}{Balanced data (IID)}\\
\toprule
FedAvg         & 0.953  $\pm$ 0.037       & 0.838  $\pm$ 0.027       & 0.764  $\pm$ 0.022  &   0.784  $\pm$ 0.012  &   0.760 $\pm$ 0.092  \\
FLTrust        & 0.952  $\pm$ 0.037       & 0.837  $\pm$ 0.022       & 0.764  $\pm$ 0.025  &   0.785  $\pm$ 0.011  &   0.761 $\pm$ 0.061         \\
Krum           & 0.935  $\pm$ 0.046       & 0.822  $\pm$ 0.021       & 0.671  $\pm$ 0.029  &   0.744  $\pm$ 0.016  &   0.781 $\pm$ 0.022         \\
TrimmedMean    & 0.952  $\pm$ 0.039       & 0.835  $\pm$ 0.025       & 0.762  $\pm$ 0.018  &   0.789  $\pm$  0.009   &   0.708 $\pm$ 0.050       \\
Sentinel       & 0.951  $\pm$ 0.041       & 0.834  $\pm$ 0.025       & 0.754  $\pm$ 0.018  &  0.763  $\pm$  0.031   &   0.750 $\pm$ 0.082     \\
\toprule
\multicolumn{6}{c}{Unbalanced data (Non-IID Dirichlet[$\alpha=0.5$])}\\
\toprule
FedAvg         & 0.941 $\pm$ 0.014 & 0.836 $\pm$ 0.006 & 0.599 $\pm$ 0.044   &   0.660 $\pm$ 0.012  &   0.653 $\pm$ 0.009    \\
FLTrust        & 0.884 $\pm$ 0.045 & 0.770 $\pm$ 0.039 & 0.459 $\pm$ 0.053    &  0.573 $\pm$ 0.017  &   0.636 $\pm$ 0.013    \\
Krum           & 0.908 $\pm$ 0.009 & 0.741 $\pm$ 0.031 & 0.375 $\pm$ 0.014   &  0.547 $\pm$ 0.018   &  0.661 $\pm$ 0.010   \\
TrimmedMean    & 0.963 $\pm$ 0.009& 0.833 $\pm$ 0.011 & 0.636 $\pm$ 0.013   &  0.668 $\pm$ 0.013  &   0.661 $\pm$  0.010   \\
Sentinel       & 0.894 $\pm$ 0.072 & 0.727 $\pm$ 0.042 & 0.448 $\pm$ 0.051    & 0.666  $\pm$ 0.012   &   0.652 $\pm$ 0.035    \\
\toprule

\end{tabular}%
}
\end{table}

\begin{figure}[h]
    \centering
    \subfloat[F1-Score Under Model Poisoning Attacks on IID distributions]{%
    \begin{minipage}{\linewidth}  
        \centering 
        \includegraphics[width=\linewidth]{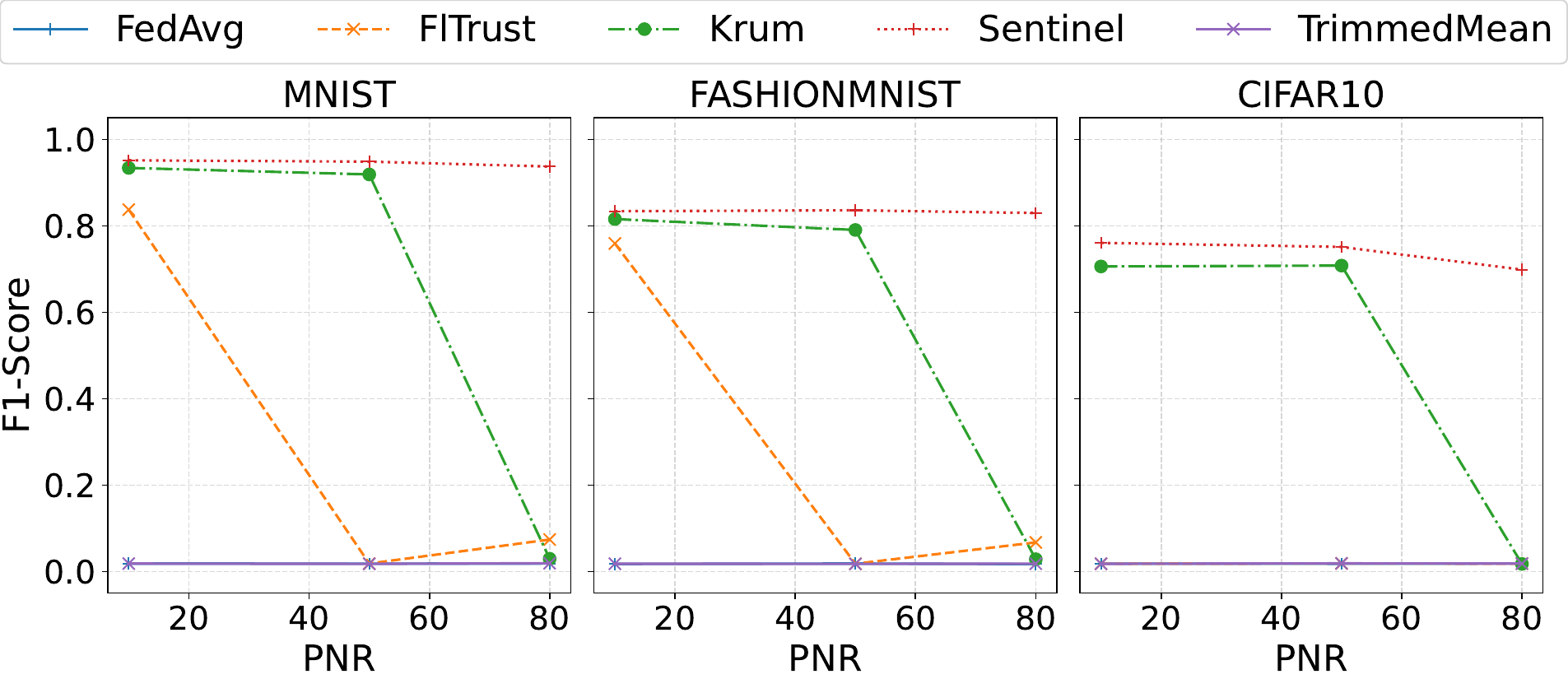}%
        \vspace{0em}
		\includegraphics[width=1\linewidth]{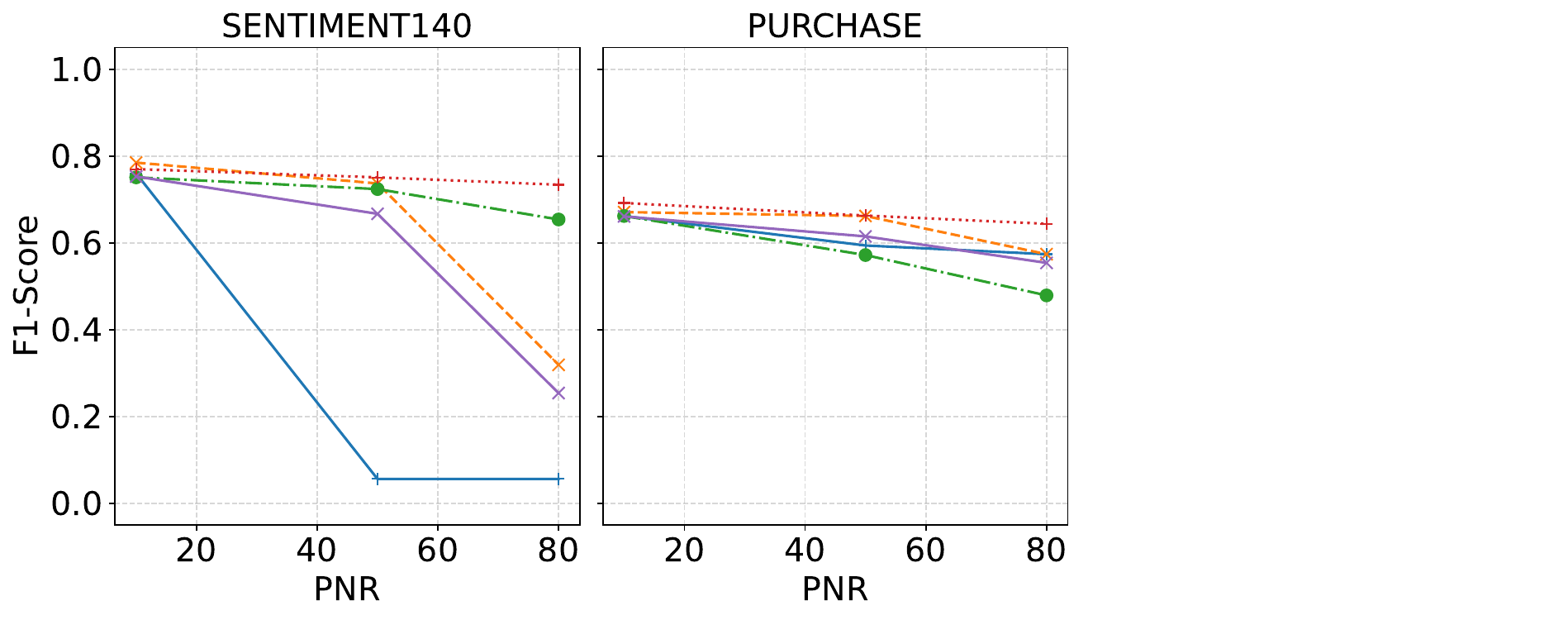}%
        \end{minipage}%
        \label{fig:model_poisoning_iid}%
    }%
     \vspace{1em}
    \subfloat[F1-Score Under Model Poisoning Attacks on non-IID distributions (Dirichlet $\alpha=0.5$)]{%
    \begin{minipage}{\linewidth}  
        \includegraphics[width=\linewidth, trim=0.25cm 0 0 0, clip]{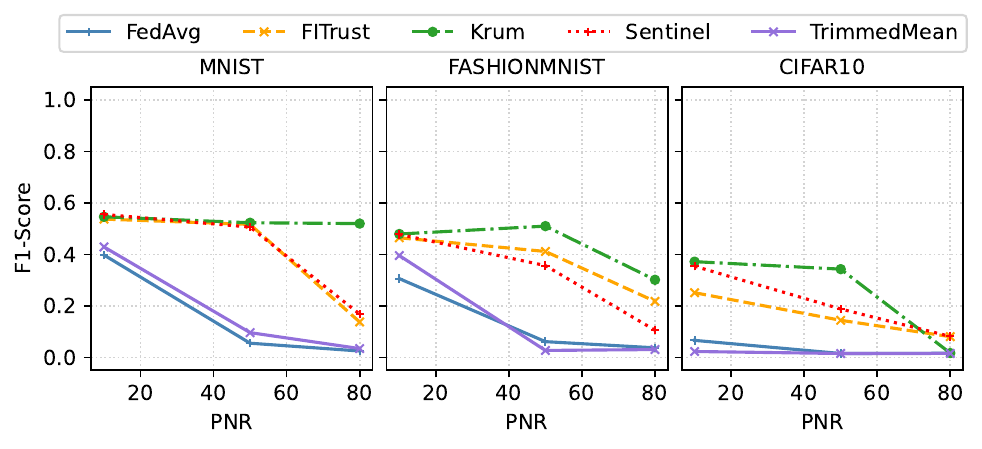}%
		  \vspace{0em}
       
		\includegraphics[width=1\linewidth]{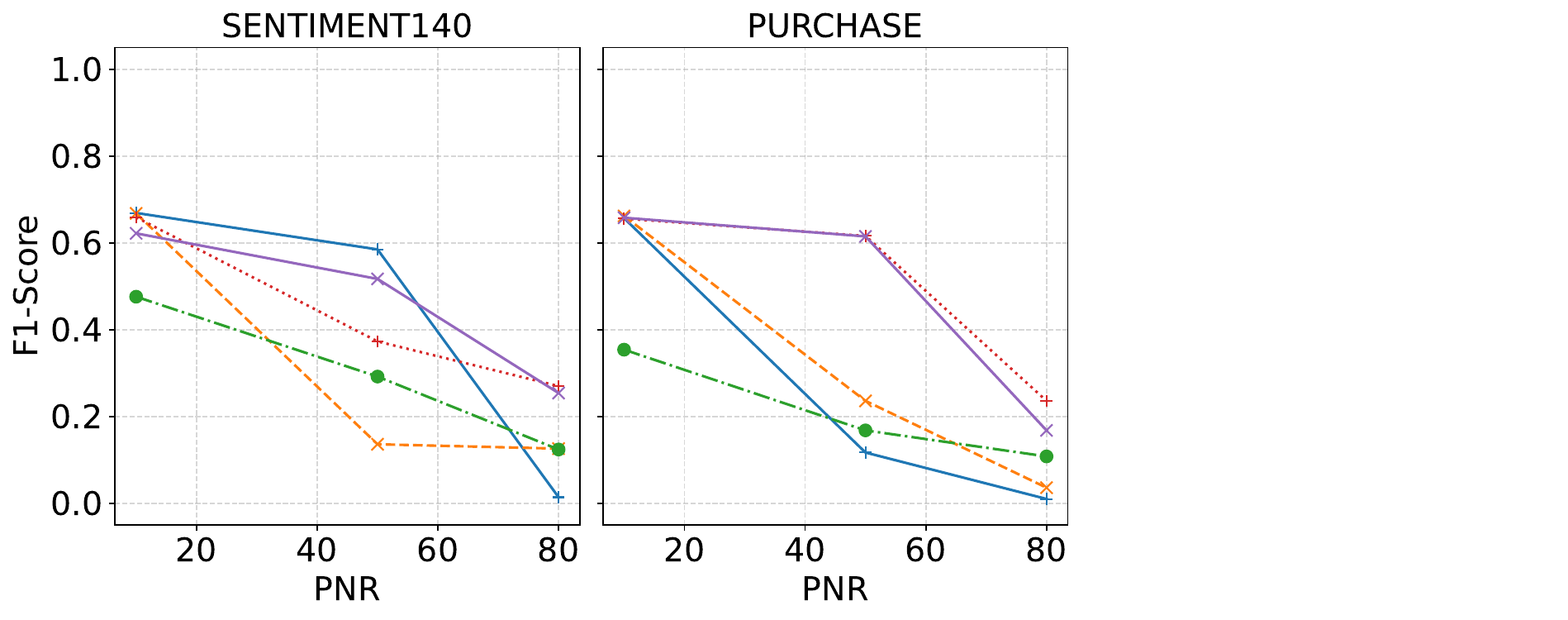}%
        \end{minipage}%
        \label{fig:model_poisoning_non_iid}%
    }
    \caption{F1-Scores Under Model Poisoning Attacks}
    \label{fig:model_poisoning}
\end{figure}

\textbf{Model Poisoning.} In this attack, the weights are randomly altered with salt noise before the model is sent to their neighbors at each round. The noise ratio (NR) remained fixed at 80\% for all configurations. Instead, the poisoned node ratio (PNR) was varied with the ratios 10\%, 50\%, and 80\% for all datasets. \figurename~\ref{fig:model_poisoning_iid} shows the F1-Score for the five datasets and each PNR configuration under IID distribution. As can be seen, Sentinel performs the best, with a maximum F1-Score reduction of $0.04$ from the baseline performance with FedAvg, observed on CIFAR10 and a PNR of 80\%. Sentinel shows the same high performance not only in the CV datasets but also in the Sentiment140 and Purchase datasets. In contrast, \figurename~\ref{fig:model_poisoning_non_iid} shows that Krum performs better when the distribution is non-IID. This is caused by a higher filtering rate from Sentinel compared to other methods, which could affect non-IID clients that are legitimate. Still, the results prove that Sentinel performance in non-IID is close to FlTrust and improves FedAvg and TrimmedMean aggregation.

\begin{figure}[h!]
    \centering
    \subfloat[F1-Score Under Untargeted Label Flipping Attacks on IID distribution]{%
    	\begin{minipage}{\linewidth}  
            \centering  
            \includegraphics[width=\linewidth]{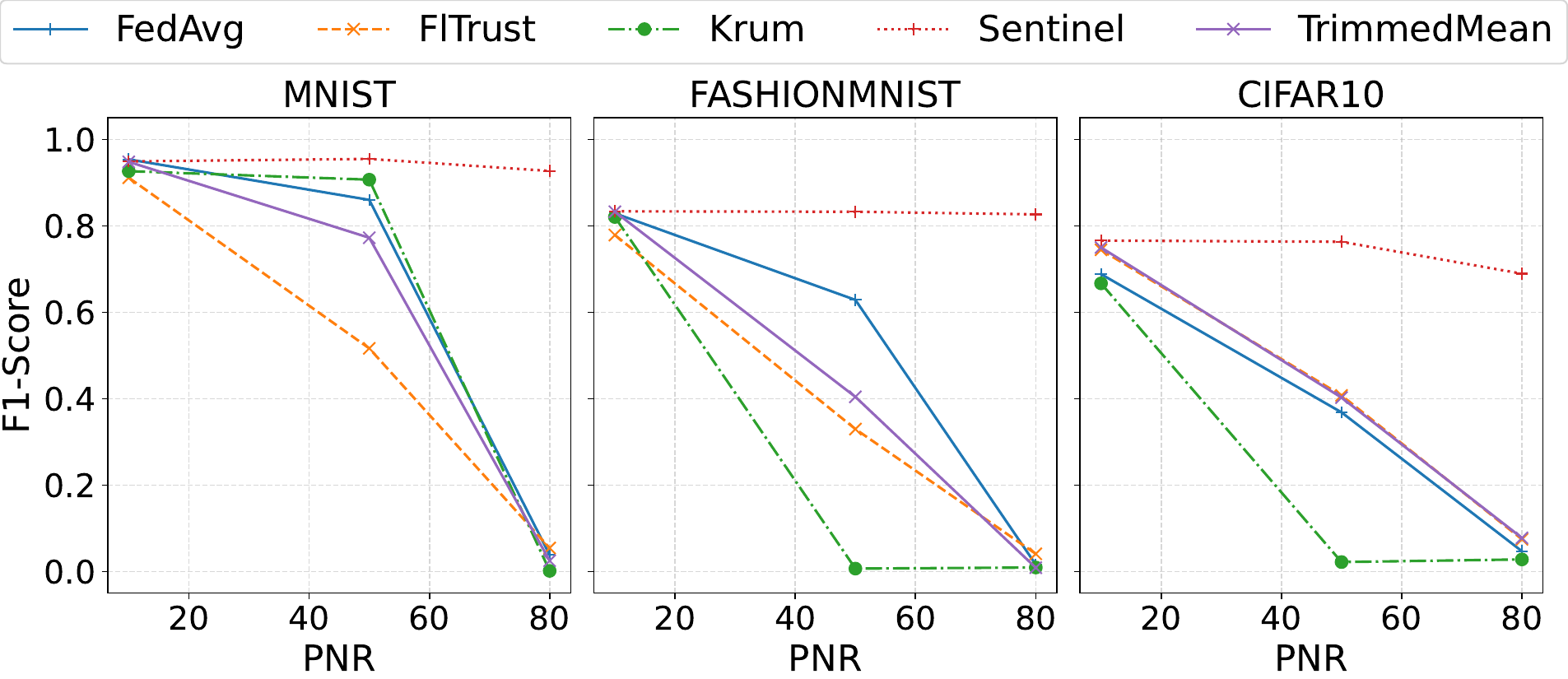}%
            \vspace{0em}
            \includegraphics[width=1\linewidth]{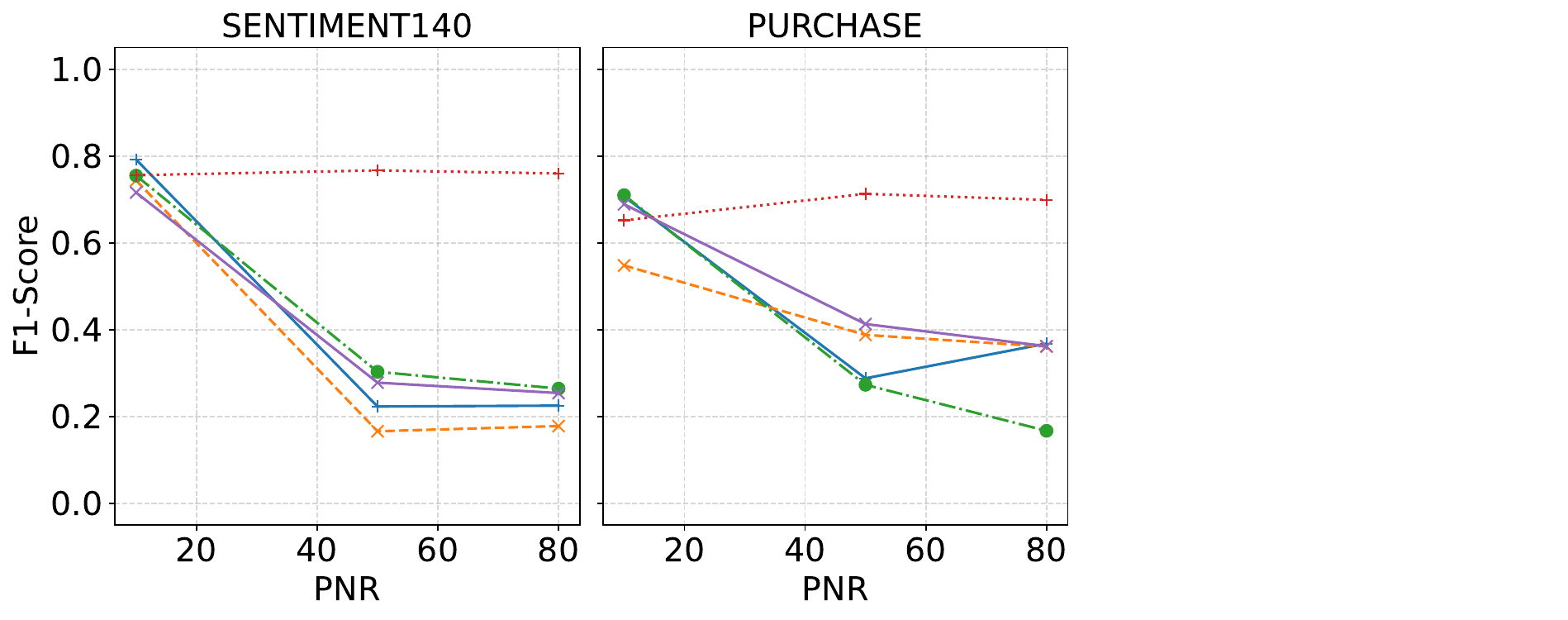}%
        \end{minipage}%
        \label{fig:untargeted_label_flipping_iid}
    }
    \vspace{1em}  
    \subfloat[F1-Score Under Untargeted Label Flipping Attacks on non-IID distribution]{%
    \begin{minipage}{\linewidth}
        \includegraphics[width=\linewidth, trim=0.25cm 0 0 0, clip]{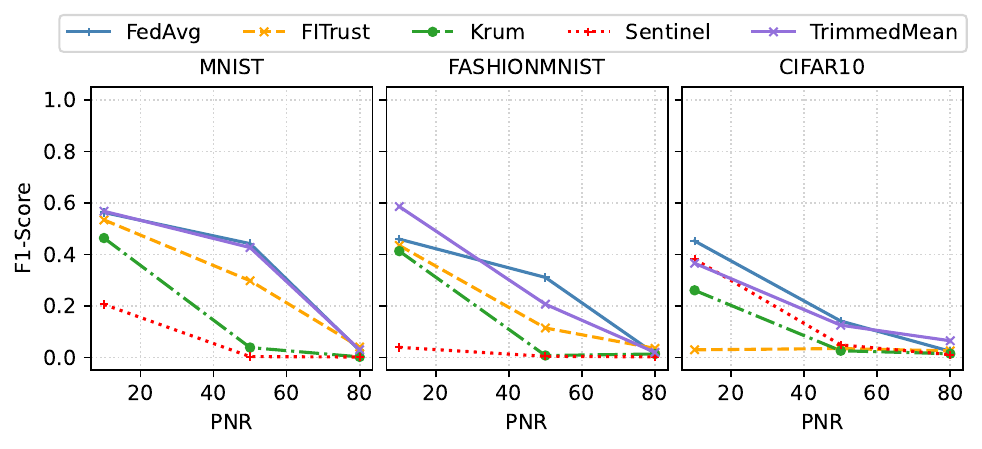}%
        \vspace{0em}
        \includegraphics[width=1\linewidth]{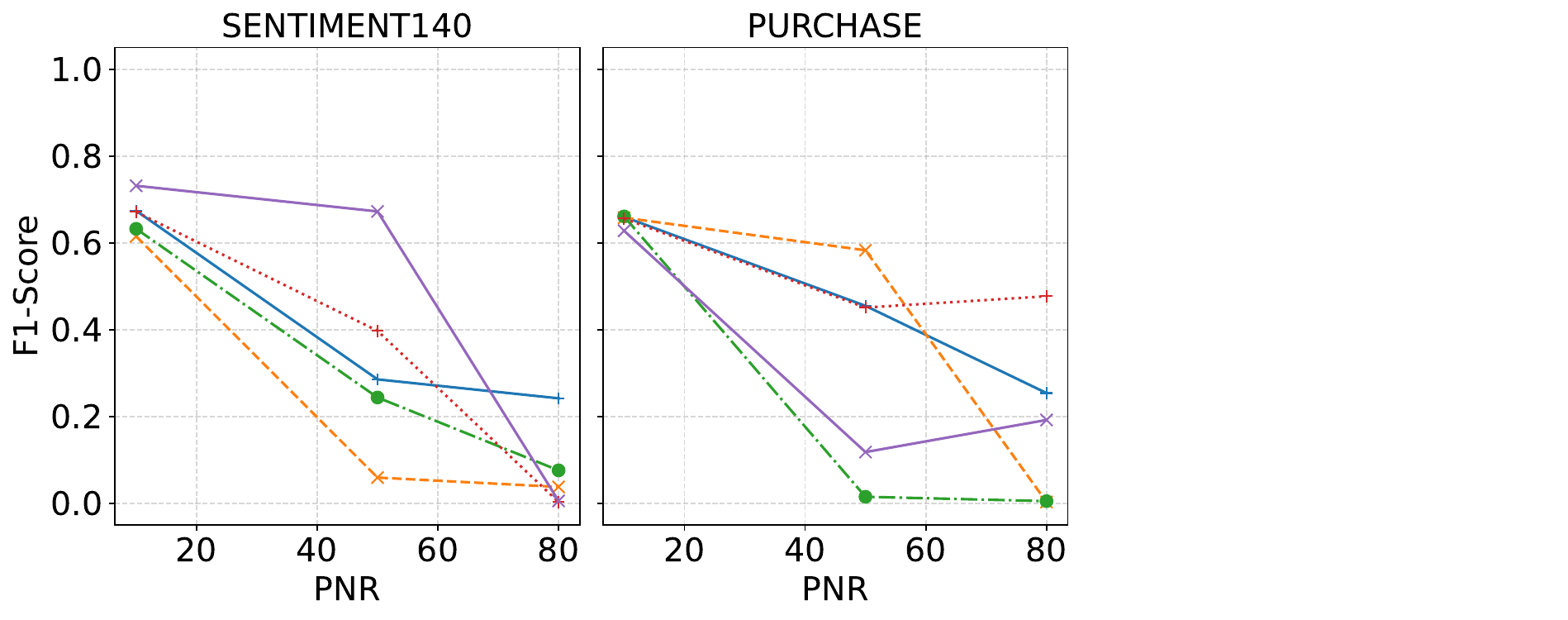}%
        \end{minipage}%
        \label{fig:untargeted_label_flipping_noniid}%
    }
    \caption{F1-Scores Under Untargeted Label Flipping Attack}
    \label{fig:label_flipping_untargeted}
\end{figure}

\textbf{Untargeted Label Flipping.} In this experiment, adversaries modify all the local training data by randomly altering the labels of their data samples. The aim is to degrade the performance of the model from benign actors, provoking random mispredictions. As can be seen in \figurename~\ref{fig:untargeted_label_flipping_iid}, Sentinel is the unique defense able to maintain its baseline F1-Score for any attack configuration and dataset when the distribution is IID. More in detail, with 80\% of malicious participants, the performance of the rest of the defense mechanisms is destroyed. However, in the case of non-IID distribution, the robust aggregation mechanisms are the ones performing the worse, especially in the case of Sentinel, as \figurename~\ref{fig:untargeted_label_flipping_noniid} shows. This can be caused by the model filtering in robust aggregation mechanisms, which could filter honest clients that generate more varied models due to their local data distribution.

To conclude, in untargeted attacks, Sentinel is effective against many types of attacks, including model poisoning and label flipping attacks. This effectiveness is well validated on different types of datasets, including CV, NLP, and tabula datasets.

\textbf{Targeted Label Flipping.} The attack causes a misclassification of a selected source label $l_{scr}$ to a target label $l_{t}$. The success of this poisoning attack is measured by the amount of test data samples labeled with the source label classified as the target label by the final models of benign participants. This ratio is represented by the Attack Success Rate for label flipping ($ASR_{lf}$), as shown in \eqref{eq:ASR_lf}, where $c_{ij}$ is the number of samples having true label $y_i$ and predicted label $\hat{y}_j$. $L$ represents the set of labels in the corresponding dataset.

This experiment was performed only in MNIST, Fashion-MNIST, and CIFAR10. These three datasets are ten classification tasks and, thus, can be attacked by targeting specific labels. However, Sentiment140 and Purchase are classification tasts, so the targeted label flipping attack is not applicable in this case.

\begin{equation} \label{eq:ASR_lf}
    ASR_{lf} = \frac{c_{src, t}}{\sum_{j=0}^{|L|} c_{src, j}}
\end{equation}

\begin{figure}[h!]
    \centering
    \subfloat[ASR Under Targeted Label Flipping Attacks on IID distribution]{%
        \includegraphics[width=\linewidth]{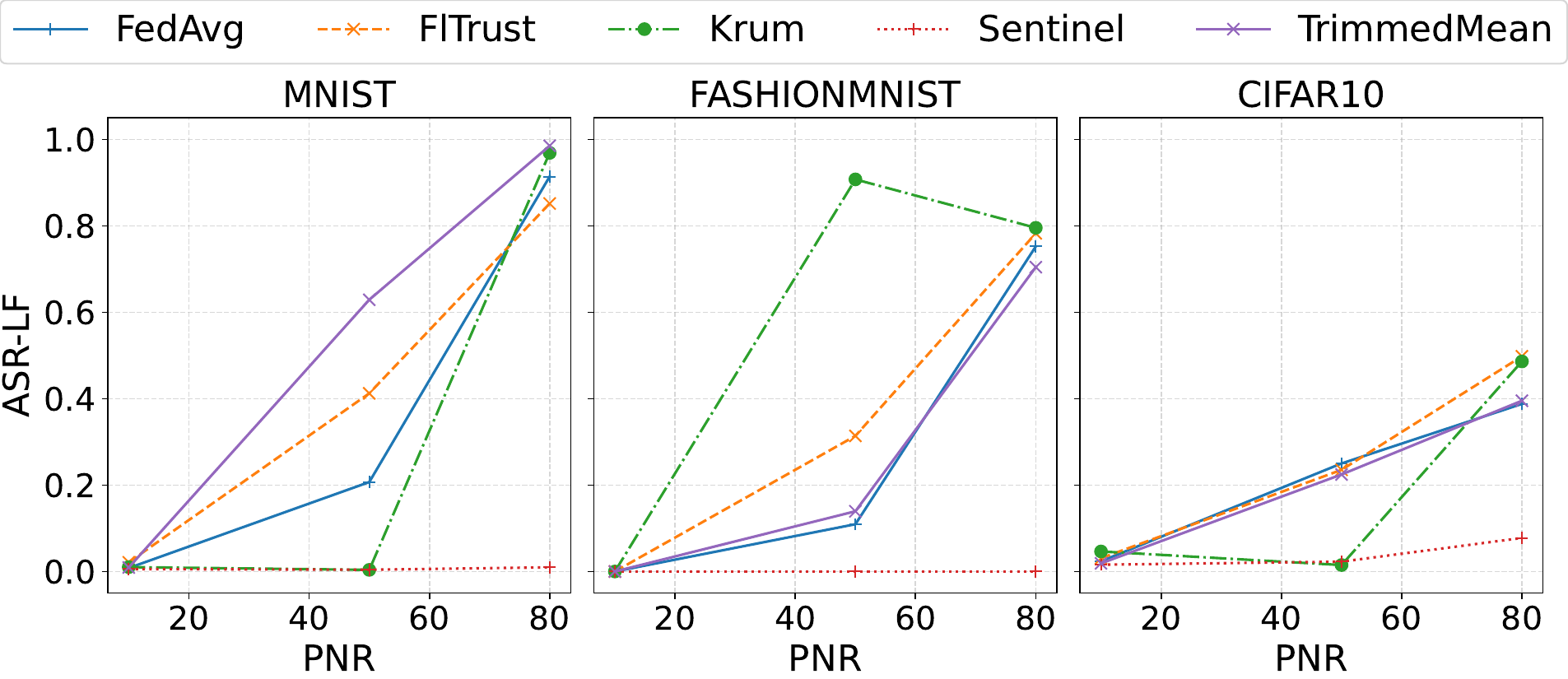}%
        \label{fig:target_label_flipping_iid}%
    }
    \hfill  
    \subfloat[ASR Under Targeted Label Flipping Attacks on non-IID distribution]{%
        \includegraphics[width=\linewidth, trim=0.25cm 0 0 0, clip]{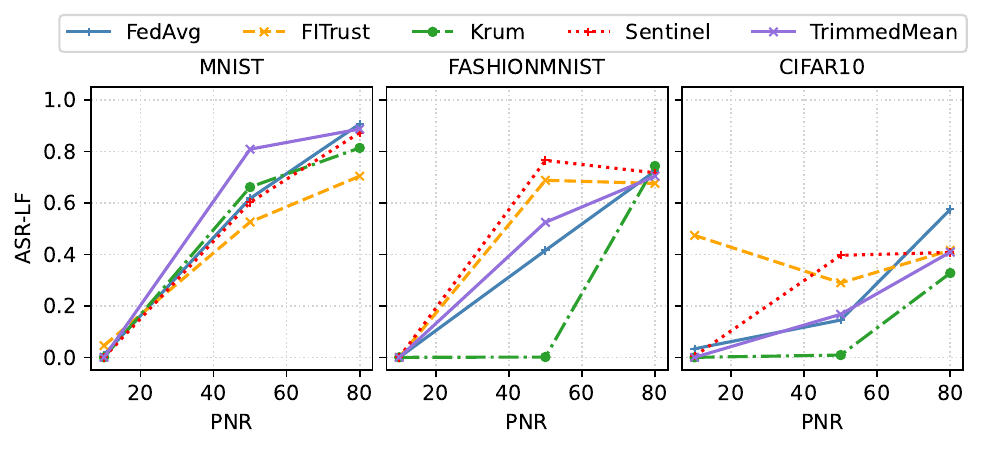}%
        \label{fig:target_label_flipping_noniid}%
    }
    \caption{ASR-LF (Attack Success Rate - Label Flipping) Under Targeted Label Flipping Attack}
    \label{fig:target_label_flipping_combined}
\end{figure}

As can be seen in \figurename~\ref{fig:target_label_flipping_iid}, Sentinel performs the best for all datasets and PNR when the distribution is IID. This can be explained by the exclusion of untrusted nodes and thereby preventing an aggregation of malicious, but highly similar, poisoned models. However, as \figurename~\ref{fig:target_label_flipping_noniid} shows, and similarly to the Untargeted setup, the robust aggregation methods do not perform well when the distribution is non-IID, again especially for Sentinel. The reason behind this lack of performance is the extensive filtering of different models based on distance, which can degrade performance under non-IID. Note that this problem is present in all robust aggregation mechanisms, not only for Sentinel. 

\textbf{Backdoor.} In this attack, malicious participants try to provoke a misclassification to a predefined target label $l_{t}$ by using an artificial trigger. In this work, this trigger is represented by the form of an “X” on any image of the three CV datasets (MNIST, Fashion-MNIST, and CIFAR10). In Sentiment140, 20\% of the training data with a sentiment of positive are randomly selected, and their first three-word embeddings are changed to 0 vectors. In Purchase, 20\% of the data in labeling '1' are selected and their first seven dimensions of the data are changed to 1.

\begin{figure}[h!]
    \centering
    \subfloat[BA Under Backdoor Attacks on IID distribution]{%
    \begin{minipage}{\linewidth}  
        \centering 
        \includegraphics[width=\linewidth]{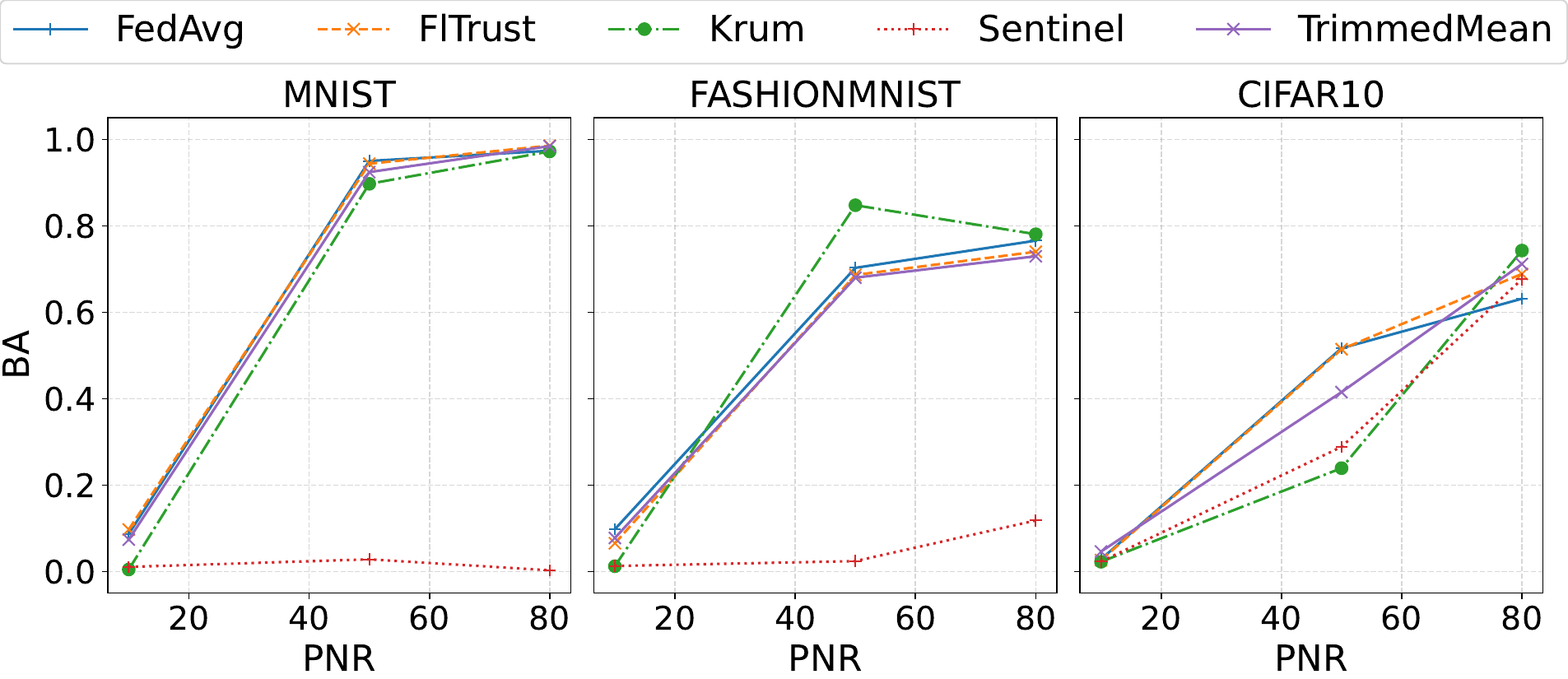}%
         \vspace{0em}
        \includegraphics[width=1\linewidth]{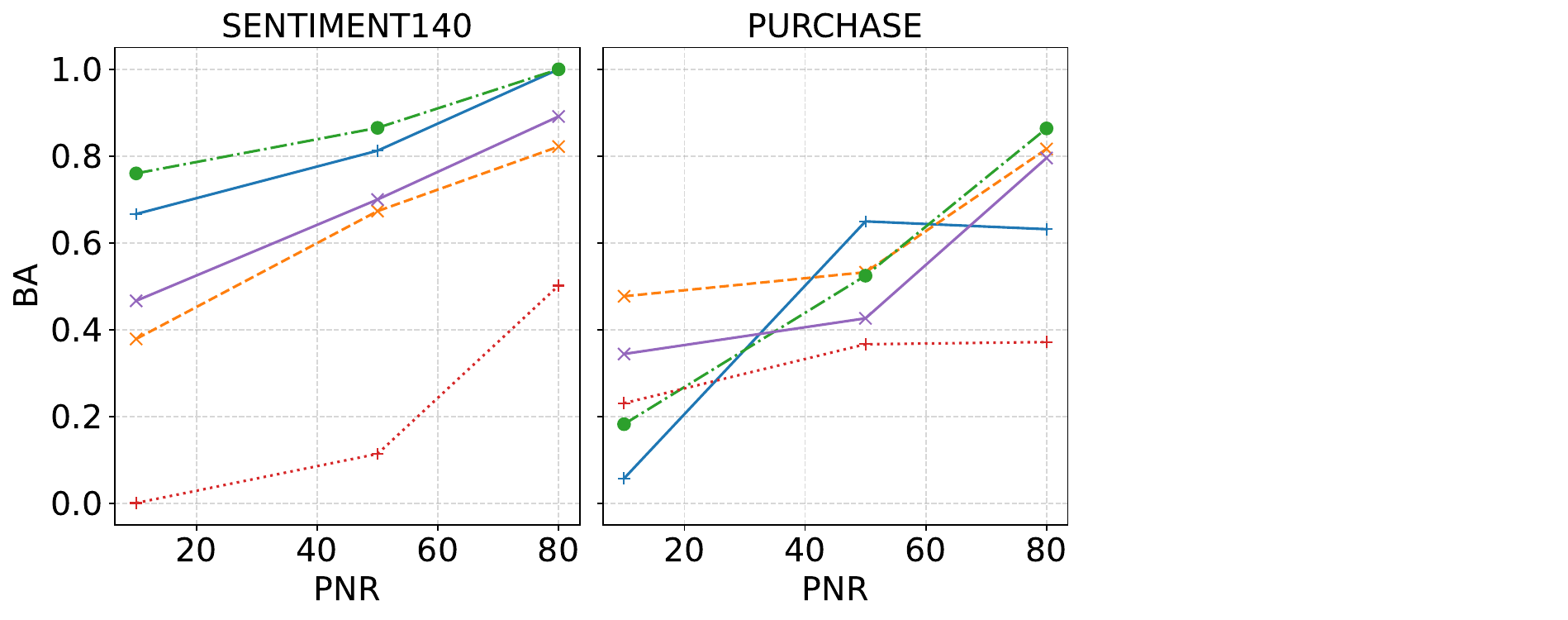}%
        \end{minipage}%
        \label{fig:backdoor_iid}%
    }
     \vspace{1em}  
    \subfloat[BA Under Backdoor Attacks on non-IID distribution]{%
    \begin{minipage}{\linewidth}  
        \centering 
        \includegraphics[width=\linewidth, trim=0.25cm 0 0 0, clip]{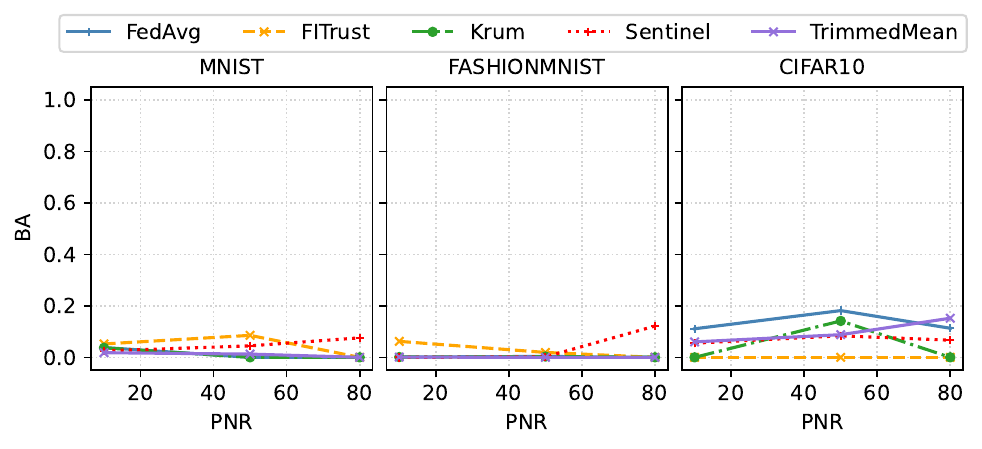}%
        \vspace{0em}
        \includegraphics[width=1\linewidth]{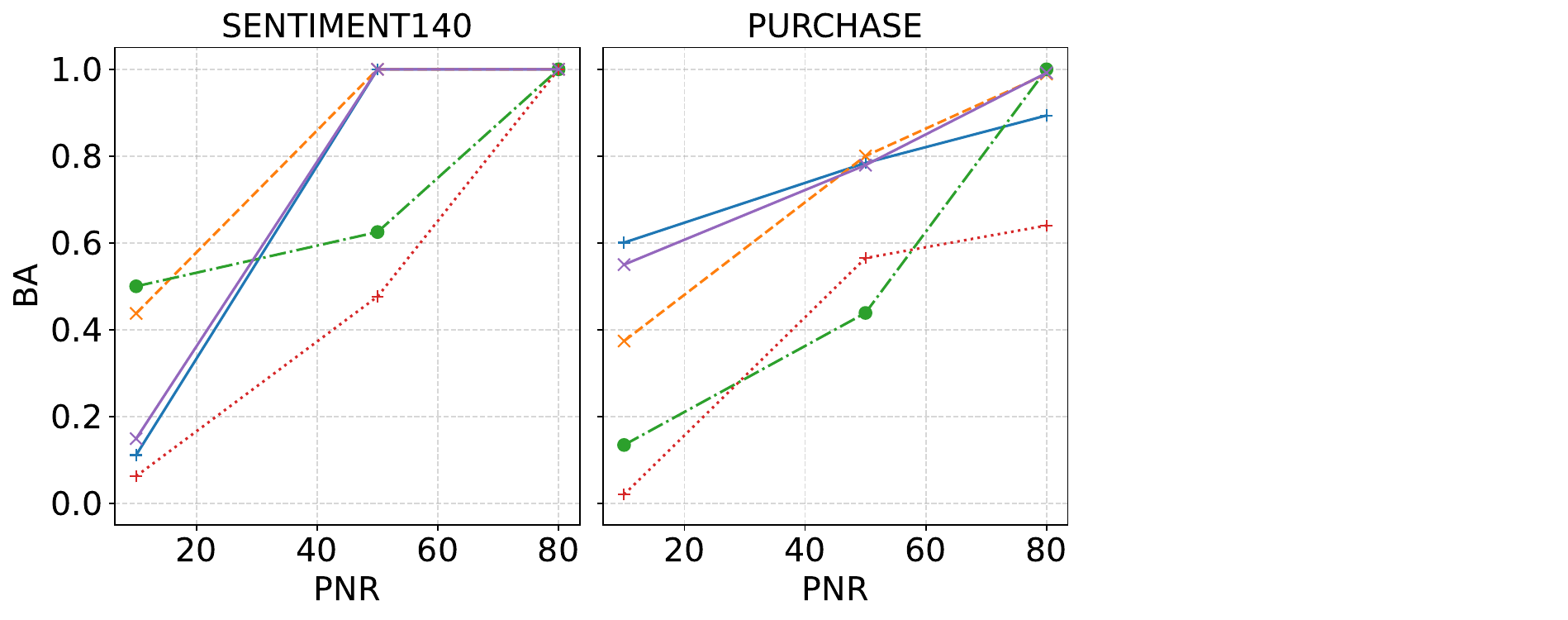}%
        \end{minipage}%
        \label{fig:backdoor_noniid}%
    }
    \caption{BA (Backdoor Accuracy) Under Backdoor Attacks on Different Distributions}
    \label{fig:backdoor_attacks_combined}
\end{figure}

Backdoor Accuracy ($BA$) is used for evaluation of defenses, as shown in \eqref{eq:BA}, where $c_{ij}$ is the number of samples having true label $y_i$ and predicted label $\hat{y}_j$. $L$ represents the set of labels in the corresponding dataset, and $B$ is the backdoor dataset.

\begin{equation} \label{eq:BA}
    BA = \frac{\sum_{j=0}^{|L|} c_{j, t} - c_{t, t}}{|B| - c_{t, t}}
\end{equation}
As can be seen in \figurename~\ref{fig:backdoor_iid}, Sentinel is the most robust solution when the distribution is IID, performing almost perfectly for MNIST, Fashion-MNIST, Sentiment140, and Purchase (regardless of the PNR). In the CIFAR10 case, Sentinel achieves similar performance to the state of the art. In the case of non-IID, the results in \figurename~\ref{fig:backdoor_noniid} show that the backdoor attack is not effective in the non-IID setup, probably due to the variations in the distributions in each local client dataset.

\section{Discussion}
\label{sec:discussion}

As experiments demonstrated, the resilience of Sentinel against a range of attacks in IID scenarios improved state-of-the-art methods. Therefore, Sentinel is a reliable defense mechanism in controlled environments. However, the fluctuating performance in non-IID settings suggests that further refinements are needed to enhance its adaptability and discrimination capabilities in more complex data distributions.

In IID, Sentinel, which sequentially integrates three distinct defensive mechanisms, effectively filters out malicious updates before aggregation. This multi-phase approach minimizes the impact of poisoning and ensures that only trustworthy model updates contribute to the aggregated global model. The experimental results validate that Sentinel can maintain integrity and performance even under aggressive attack scenarios, such as model poisoning and targeted label flipping.

While Sentinel effectively secures federated learning networks in IID scenarios by filtering out anomalous updates, its performance under non-IID conditions requires careful consideration. The distinct data characteristics inherent to non-IID distributions can lead to high variability in local model updates. This variability could be misinterpreted by Sentinel similarity filtering phase as potentially malicious, leading to the unwarranted exclusion of legitimate model updates. Such exclusions reduce the overall model quality and affect learning efficiency and convergence rates.

Furthermore, the bootstrap validation step in Sentinel, which assesses model updates based on their loss performance on a small, locally held validation dataset, can penalize nodes that genuinely represent minority patterns within the data. Since these nodes naturally exhibit higher loss values due to their divergent data characteristics, their contributions might be undervalued in the aggregation process. This scenario underscores a critical trade-off in Sentinel design between robustness to attacks and sensitivity to the legitimate diversity of participant data.

To better accommodate non-IID distributions, future iterations of Sentinel could benefit from incorporating adaptive mechanisms that adjust the thresholds for similarity and loss validation based on the observed data distribution variance among nodes. For instance, dynamic adjustment of the similarity threshold ($\tau_{S}$) and the loss distance threshold ($\tau_{L}$) could help mitigate the risk of excluding non-malicious yet statistically unusual data patterns. Additionally, implementing anomaly detection techniques that consider the context of data distribution or employing unsupervised learning to better understand the typical behavior of node updates could enhance Sentinel capability to discern between attacks and naturally occurring data anomalies.

\section{Conclusion and Future Work}
\label{sec:conclusion}

This work has proposed Sentinel, a sophisticated defense strategy that applies a multi-level defense protocol composed of similarity filtering, bootstrap validation, and model normalization to mitigate poisoning attacks. By taking advantage of the local data availability in DFL, Sentinel demonstrates the transferability of promising defense strategies proposed for CFL. Extensive evaluations on the datasets MNIST, Fashion-MNIST, CIFAR10, Sentiment140, and Purchase datasets with various attack configurations have demonstrated their effectiveness. Compared to other state-of-the-art aggregation algorithms, the aggregation protocol proposed in this work reliably defended against poisoning attacks under IID configurations, especially in highly malicious environments. Furthermore, this work demonstrated the usability of Fedstellar for more comparable security benchmarks in DFL. In contrast, under non-IID configurations Sentinel shows some disadvantages compared to traditional non-robust aggregation methods, drawbacks that are also present in other robust methods.

As future work, it is planned to optimize the Sentinel algorithm better to handle the variability and unpredictability of non-IID data. Exploring hybrid models that combine Sentinel robust aggregation with other techniques could potentially yield methods that are robust to attacks and sensitive to legitimate data diversity. Other relevant directions are evaluating scenarios involving alternative network topologies, such as ring or random graphs, with or without clustered components. Finally, evaluating the impact of different Dirichlet values in the non-IID is another experiment for the future. 

\section*{Acknowledgments}

This work has been partially supported by \textit{(a)} the Swiss Federal Office for Defense Procurement (armasuisse) with the CyberForce (CYD-C-2020003) project, \textit{(b)} the University of Zürich UZH, \textit{(c)} 21629/FPI/21, Fundación Séneca, Región de Murcia (Spain), and \textit{(d)} the strategic project DEFENDER from the Spanish National Institute of Cybersecurity (INCIBE) and by the Recovery, Transformation and Resilience Plan, Next Generation EU.

\bibliography{references}

\end{document}